\documentclass[final,5p,times,twocolumn]{elsarticle}
 
\usepackage{longtable}
\usepackage{cancel}
\usepackage{amssymb,mathrsfs}
\usepackage{bbm,bm}
\usepackage{amsmath}
\usepackage{siunitx}
\usepackage{braket}
\usepackage{mathtools}
\usepackage [utf8] {inputenc}
\usepackage{color}
\usepackage{lmodern}

\usepackage{footnote}
\usepackage{slashed}
\usepackage{mathrsfs}
\usepackage{hyperref}
\usepackage[list=true, labelfont=bf, labelformat=brace, position=top]{subcaption}
\usepackage[dvipsnames,svgnames,table]{xcolor}
\usepackage{lipsum}

\newcommand{\eq}{\begin{eqnarray}}
\newcommand{\en}{\end{eqnarray}}

\journal{Physics Letters B}

\begin{document}

\begin{frontmatter}



\title{Effective Field Theory of Protonium}


\author[first1,first2]{Hans-Werner Hammer}
\affiliation[first1]{
            addressline={Technische Universität Darmstadt, Department of Physics, Institut f\"ur Kernphysik, 64289 Darmstadt}, 
            country={Germany}}
\affiliation[first2]{
            addressline={ExtreMe Matter Institute EMMI and Helmholtz Forschungsakademie Hessen für FAIR (HFHF), GSI Helmholtzzentrum für Schwerionenforschung GmbH, 64291 Darmstadt},
            country={Germany}}

            \author[second1]{Akaki Rusetsky}
\affiliation[second1]{
            addressline={Helmholtz-Institut f\"ur Strahlen- und Kernphysik (Theorie) and Bethe Center for Theoretical Physics, Universit\"at Bonn, 53115 Bonn}, 
            country={Germany}}

\begin{abstract}
We investigate the level shifts in protonium using the framework of non-relativistic effective field theory (NREFT). Our study is prompted by the PUMA collaboration's plans to probe the neutron-to-proton ratio in the nuclear density tail using antiprotonic atoms -- a method for which light atoms provide an important benchmark. We calculate the corrections  to the Deser-Goldberger-Baumann-Thirring formula for the
complex level shift  of S-wave states from the unitary cusp and Coulomb photon exchange. Structure dependent corrections enter at the next order. For higher-partial-wave states, they enter  already at leading order. Using input from chiral $N\bar N$ interactions, we compare to other calculations and measurements from LEAR.
\end{abstract}




\end{frontmatter}




{\bf 1.} The antiProton Unstable Matter Annihilation
(PUMA) experiment at CERN \cite{PUMA:2022ngr} will use low-energy antiprotons as a new method to probe the neutron-to-proton ratio in
the nuclear density tail, where the neutron skin of heavy nuclei develops. 
The ratio of neutron-antiproton to proton-antiproton annihilations following the antiproton capture on a nucleus to form an antiprotonic atom will be determined from the detection of the charged pions produced in the annihilations. This ratio is connected to the ratio of neutron and proton densities at the annihilation site. Quantifying this connection in a systematic way is the subject of ongoing investigations.

While the main focus of PUMA is on heavy nuclei which display a neutron skin, measurements on stable light nuclei,
such as hydrogen and helium isotopes, are used to benchmark both the experimental setup and method, as well as to validate the theoretical approach. The simplest antiprotonic atom is protonium, consisting of an antiproton bound to a proton. Protonium has been studied experimentally and theoretically  over more than six decades \cite{Desai:1960zz,Buck:1977rt,Carbonell:1989cs,Carbonell:1992wd,Vandermeulen:1992eh,Kong:1998ps,Klempt:2002ap,Ydrefors:2021pmp,Cabrera-Trujillo:2023ify,Ma:2024gsw}, highlighting its role as a unique laboratory for testing low-energy strong interactions. Early foundational work established the basic framework for proton-antiproton annihilation dynamics and statistical phase-space decays \cite{Desai:1960zz,Vandermeulen:1992eh}. 
Moreover, coupled channel models have been used to calculate the annihilation densities \cite{Ydrefors:2021pmp}.
For review articles on protonium, see refs.~\cite{Klempt:2002ap,Ma:2024gsw}.

More recently, the integration of non-relativistic effective field theory and combined QCD plus QED frameworks allowed researchers to calculate higher-order hadronic energy level shifts in a systematic way. For instance, the pioneering study of protonium within this approach was carried out in Ref.~\cite{Kong:1998ps}. General reviews of the method are given in Refs~\cite{Gasser:2007zt,Gasser:2009wf}. 
In this work, we perform NREFT analysis of protonium and  calculate the level shifts. A particular focus is placed on the
unitary cusp and Coulomb corrections which turn out to be sizable.

{\bf 2.}
As already mentioned above,
the spectrum and decays of many hadronic atoms have been systematically studied within
the framework of the non-relativistic effective field theory (NRFET).
The characteristic energy scale in the $p\bar p$ atom is given by the ground-state binding energy
$E_0=\frac{1}{4}\,m\alpha^2\simeq 12.5\,\mbox{keV}$
(here, $m$ denotes the proton mass and $\alpha$ is the fine structure constant).
This is still an order of magnitude larger than the strong shift of the ground-state level
and much less than the proton-neutron mass difference. Furthermore, the characteristic length scale is given by the Bohr radius of the atom,
$a_0=\frac{2}{m\alpha}\simeq 58\,\mbox{fm}$, which is substantially larger than the proton radius, $r_p\simeq 1 \,\mbox{fm}$, or the nucleon-antinucleon scattering lengths,
$|a_{p\bar p}| \simeq 1-2\,\mbox{fm}$. From this one may
conclude that $p\bar p$ atom should be amenable for study in NREFT, albeit the scale hierarchy is not as pronounced as in the case of similar systems, like pionium or pionic hydrogen. The corrections might be more sizable and should be taken into account. A previous NREFT study of protonium by Kong and Ravndal \cite{Kong:1998ps} was limited to the ground state and did not consider corrections from Coulomb interactions and the unitary cusp, which turn out to be very important. In this
short note, we address these issues.

The energy shift and the width of any atomic level are determined by the real and imaginary parts of the complex energy shift, where the latter is connected to the lifetime of the atom. Their calculation will be presented
below. To start with, it is conventional to split this complex shift 
into the ``electromagnetic'' and ``strong'' parts. The calculation of the ``electromagnetic''
part is carried out in pure QED and includes contributions from the transverse photon
exchanges, electron vacuum polarization and the electromagnetic radius of the nucleon.
These calculations are pretty standard (see, e.g., Ref.~\cite{Gasser:2007zt}), and we do not consider them here, assuming
that they are done elsewhere and are already subtracted from the experimentally
determined value of the level shift. Thus, we concentrate below on the strong shift only.
However, before performing the calculations, we wish to discuss the nature of the
corrections to the leading-order strong shift and to quantify the systematic error in these
calculations. It should be emphasized that the effective field theories provide a unique
tool to achieve this goal.

In the NREFT, the nucleons and antinucleons, interacting with photons, are described by the
non-relativistic Lagrangian
    \eq\label{eq:Lfull} 
\mathscr{L}
&=&\psi^\dagger\left(i\partial_t-m+\frac{\Delta}{2m}+\cdots\right)\psi
\nonumber\\
&+&\psi_A^\dagger\left(i\partial_t-m+\frac{\Delta}{2m}+\cdots\right)\psi_A
\nonumber\\
&+&\chi^\dagger\left(i\partial_t-m_n+\frac{\Delta}{2m_n}+\cdots\right)\chi
\nonumber\\
&+&\chi_A^\dagger\left(i\partial_t-m_n+\frac{\Delta}{2m_n}+\cdots\right)\chi_A
\nonumber\\
&+&\mathscr{L}_C+\mathscr{L}_S+\cdots\, ,
\en
where $\Delta$ is the Laplace operator.
Moreover, $\psi$, $\psi_A$, $\chi$, and $\chi_A$ are the proton, antiproton, neutron, and antineutron
fields, respectively, $m$, $m_n$ are the corresponding masses, and
$\mathscr{L}_C$ describes the Coulomb force between the proton and the antiproton.
$\mathscr{L}_C$ is non-local and can be written as
\eq
\mathscr{L}_C=-e^2(\psi_A^\dagger\psi^\dagger)\Delta^{-1}(\psi\psi_A)\, .
\en
Furthermore, $\mathscr{L}_S$  denotes the strong Lagrangian
\eq
\mathscr{L}_S&=&-\psi_A^\dagger\psi^\dagger H^{(1)}\psi\psi_A
-\left(\psi_A^\dagger\psi^\dagger H^{(2)}\chi\chi_A+\mbox{h.c.}\right)
\nonumber\\
&-&\chi_A^\dagger\chi^\dagger H^{(3)}\chi\chi_A\, .
\label{eq:lag_strong}
\en
Here, the $H^{(i)}$ are local differential operators which are most conveniently written down in momentum space, 
see e.g., Ref.~\cite{Epelbaum:2004fk}.
Because of Galilei invariance, they depend only on the relative three-momenta $\bm{p}$ and $\bm{p}'$ of the incoming and outgoing nucleons,
\eq
&&\langle \bm{p}'|H^{(i)}|\bm{p}\rangle
=V_0^{(i)}+V_\sigma^{(i)}\boldsymbol{\sigma}_1\cdot\boldsymbol{\sigma}_2
\nonumber\\
&&~~+\,\frac{i}{2}\,V_{SL}^{(i)}(\boldsymbol{\sigma}_1+\boldsymbol{\sigma}_2)\cdot
(\bm{k}\times\bm{q})
\nonumber\\
&&~~+\,V_{\sigma L}^{(i)}(\boldsymbol{\sigma}_1\cdot (\bm{q}\times\bm{k}))\,
(\boldsymbol{\sigma}_2\cdot (\bm{q}\times\bm{k}))
\nonumber\\
&&~~+\,V_{\sigma q}^{(i)}(\boldsymbol{\sigma}_1\cdot \bm{q})(\boldsymbol{\sigma}_2\cdot \bm{q})
+V_{\sigma k}^{(i)}(\boldsymbol{\sigma}_1\cdot \bm{k})(\boldsymbol{\sigma}_2\cdot \bm{k})\, ,
\en
where $\bm{q}=\bm{p}'-\bm{p}$ and $\bm{k}=\frac{1}{2}\,(\bm{p}'+\bm{p})$. Because the strong interaction is purely short-range at the energy scales of protonium,\footnote{Since the non-analyticities of pion exchange are not resolved, there is no need to include pion degrees of freedom explicitly.} the functions $V_\alpha^{(i)}$ are polynomials in the momenta $\bm{q},\bm{k}$. The expansion coefficients are low-energy constants (LECs)
of the $\bar NN$ interaction. In contrast to the $NN$ interactions, these LECs are complex and their imaginary parts encode the strong annihilation effects. At the accuracy of our calculation, only the momentum-independent terms $V_0$ and $V_\sigma$ contribute. They can be related to the spin-singlet and spin triplet $p\bar p$ scattering lengths. For simplicity, we do not give the explicit expressions here. In addition, the ellipses in Eq.~(\ref{eq:Lfull}) stand for terms featuring transverse photons, non-minimal photon-nucleon-antinucleon couplings, etc. These terms do not contribute at the accuracy we are working at.

\begin{figure}[t]
  \begin{center}
    \includegraphics[width=8.cm]{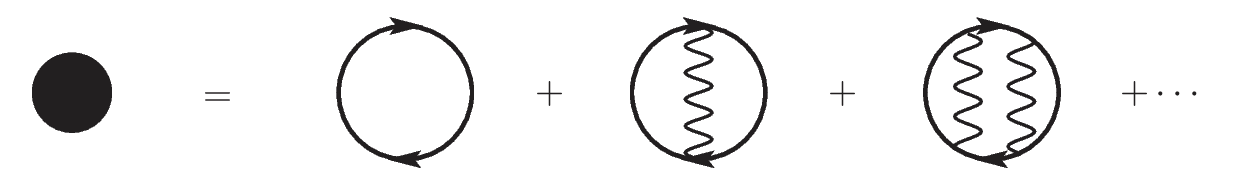}
    \caption{The filled circle represents the sum of ``Coulomb bubbles'', i.e., the bubble diagrams that contain
    the exchange of $0,1,2,\ldots$ Coulomb photons.}
    \label{fig:Coulomb_bubble}
  \end{center}
  \end{figure}

{\bf 3.} In QCD plus QED, the fine structure constant $\alpha=e^2/(4\pi)$ and the
up/down quark mass difference $m_d-m_u$ parametrize the corrections to the
leading-order Deser-Goldberger-Baumann-Thirring (DGBT) formula for the level shift \cite{Deser:1954vq}.\footnote{These parameters are of course intertwined, since the RG running of the quark masses depends on $\alpha$. Hence, the splitting can be carried out after fixing the scale only, see e.g., Ref.~\cite{Gasser:2003hk} for more discussion of this issue.} Namely, these corrections can be expanded in powers of $\alpha$ and $m_d-m_u$ (modulo logarithms). Since the neutron-proton mass difference $m_n-m$ contains both parameters linearly at the lowest order, it is conventional as well as convenient to introduce a generic small parameter $\epsilon\sim\alpha\sim (m_d-m_u)/\Lambda_{\rm had}$, where $\Lambda_{\rm had}$ is the typical hadronic scale, and expand the result in powers of $\epsilon$. We shall see below that two leading terms in the expansion are parameter-free and can be reliably calculated. The subleading terms define the systematic error. Calculating these terms is beyond the scope of this work.

The leading-order strong shift in the S-wave state is of order $\epsilon^3$
and is given by the DGBT formula \cite{Deser:1954vq},
\eq\label{eq:DeltaEDGBT}
\Delta E_1=|\tilde\Psi_1(0)|^2\frac{4\pi a_{p\bar p}}{m}=\frac{1}{2}\,\alpha^3m^2 a_{p\bar p}\, ,
\en
where $\tilde\Psi_1(0)$ is the coordinate space wave function of the protonium ground state at the origin and 
$a_{p\bar p}$ is the strong $p\bar{p}$ scattering length.\footnote{The overall sign in Eq.~(\ref{eq:DeltaEDGBT}) depends on the sign convention for the scattering length $a_{p\bar p}$. We use the convention, in which $\mbox{Im}\,a_{p\bar p}<0$.}  The leading correction to this result in the
S-wave is of order $\epsilon^{1/2}$ and comes from the effect of the unitary cusp associated with the $n\bar n$ intermediate state \cite{Gasser:2007zt}. Other intermediate states, both below and above
$p\bar p$ threshold, are distant and their contribution is analytic in $\epsilon$. Thus, the
corrections emerging from these states start at $O(\epsilon)$. Another source of potentially large corrections comes from the multiple Coulomb photon exchange diagrams shown
in Fig.~\ref{fig:Coulomb_bubble}. This correction scales as $\alpha\ln\alpha$ and not
as just $\alpha$. The presence of a factor $\ln\alpha\simeq -5$ renders this correction
parametrically enhanced, as compared to other $O(\epsilon)$ contributions.

It is important to emphasize that both $O(\epsilon^{1/2})$ and $O(\epsilon\ln\epsilon)$ corrections for the S-wave bound do not contain additional parameters and can be expressed in terms of the $N\bar N$ scattering lengths in the ``purely strong'' isospin-symmetric world. This statement
ceases to be valid for the corrections of order $\epsilon$. First note that the
LECs in the Lagrangian~(\ref{eq:Lfull}) implicitly depend both $m_d-m_u$ and $\alpha$.
For simple systems, like $\pi\pi$, $\pi K$ and $\pi N$ atoms, it was possible to estimate
$O(\epsilon)$ contributions by matching to the one-loop ChPT
results~\cite{Gasser:1999vf,Gall:1999bn,Kubis:2001ij,Kubis:2001bx,Gasser:2002am}.
For the $p\bar p$ bound states, this would require 
matching to non-perturbative calculations with chiral $N\bar N$ interactions \cite{Dai:2017ont},
which are not very well known. Thus we consider $O(\epsilon)$ corrections as a part of the systematic error.

In bound states with higher orbital momentum, however, the situation is reversed. 
On the one hand, the strong shift in the partial wave with angular momentum $\ell$ scales like $\epsilon^{2\ell+3}$ \cite{Trueman:1961zza}.
However,
the correction from the unitary cusp scales as $O(\epsilon^{\ell+1/2})$, in accordance
with the threshold behavior of the phase space in the partial wave with the angular
momentum $\ell$ (see Eq.~\eqref{eq:threshold} below), whereas the Coulomb photon exchanges scale as
$O(\epsilon^{2\ell+1})$. On the other hand, the isospin-breaking corrections to the LECs
are still of order $\epsilon$ for all partial waves. Thus in all partial waves except
the S-wave, isospin-violating $O(\epsilon)$ corrections dominate and any controlled calculation beyond the leading-order DGBT formula is extremely difficult (see Ref.~\cite{Gasser:2007zt} for more details). For this reason, we concentrate on the S-wave states and, in particular, neglect S-D wave mixing in the triplet state.

Finally, note that the simple counting considered above is valid in dimensional regularization only. It is modified if one uses, for example, cutoff regularization or solves the Schr\"odinger equation numerically with strong potentials smoothly vanishing at large momenta. Put differently, the counting of the {\em non-analytic} terms in $\epsilon$ is the same, but a polynomial piece in $\epsilon$ is not excluded. This might render the picture less transparent than it actually is. In reality, we are not able to estimate these polynomial terms in a reliable manner.

{\bf 4.} After these preliminary remarks we turn directly to the evaluation of the strong shift for the S-wave bound state. We keep the presentation concise, since the calculations are pretty standard~\cite{Gasser:2007zt,Meissner:2004jr,Meissner:2006gx}. Since the neutron-proton mass difference is much larger than the atomic ground-state energy, the $n\bar n$ channel can be integrated
out. This amounts to replacing the strong $p\bar p\to p\bar p$ scattering length $a_{p\bar p}$ by
the amplitude $A_{p\bar p}$
\eq
a_{p\bar p}&\to& A_{p\bar p}=a_{p \bar p}+\frac{a_{p\bar p\to n\bar n}^2\delta}{1-a_{n\bar n}\delta}\, ,
\en
with
\eq
\delta&=&\sqrt{2m(m_n-m)}=O(\epsilon^{1/2})\, ,
\label{eq:deltadef}
\en
where $a_{p\bar p\to n\bar n}$ and $a_{n\bar n}$ denote scattering length in the $\bar pp\to \bar nn$ and $\bar nn\to\bar nn$ channels, respectively (assuming isospin symmetry). 
Remember that, at the accuracy we are working, it suffices to consider only the non-derivative
couplings in the Lagrangian \eqref{eq:lag_strong} that can be matched to the scattering lengths in the pertinent channels.

Furthermore, the Lippmann-Schwinger equation for the $p\bar p\to p\bar p$ scattering
at the order we are working can be written as
\eq
&&\hspace*{-1.cm}T(\bm{p}',\bm{p})=V(\bm{p}',\bm{p})
\nonumber\\
&&\hspace*{-0.cm}~~+\,\frac{m}{4\pi}\,
\int\frac{d^3\bm{p}''}{(2\pi)^3}\,
\frac{V(\bm{p}',\bm{p}'')}{\bm{p}^2-{\bm{p}''}^2+i\varepsilon}\, T(\bm{p}'',\bm{p})\, ,
\label{eq:Tmatrix}
\en
with the full potential
\eq
V(\bm{p}',\bm{p})=\frac{4\pi A_{p\bar p}}{m}-\frac{4\pi \alpha}{|\bm{p}'-\bm{p}|^2}\, ,
\label{eq:pot}
\en
consisting of the strong and Coulomb potentials, respectively.
Note that the strong potential in Eq.~\eqref{eq:pot} only includes the S-wave contribution.

\begin{figure}[t]
  \begin{center}
    \includegraphics[width=8.cm]{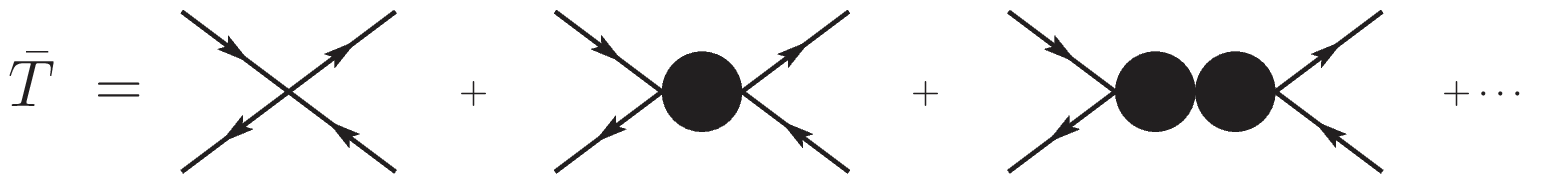}
    \caption{The amplitude $\bar T(E)$ expressed through the sum of Coulomb bubbles (filled circle) defined in Fig.~\ref{fig:Coulomb_bubble} and the strong potential. The first diagram represents the strong potential. In the further diagrams there is an insertion of the strong potential whenever external lines connect to the sum of Coulomb bubbles and when two Coulomb bubbles connect.}
    \label{fig:Tbar}
  \end{center}
  \end{figure}

The energy shift of the S-wave state with the principal quantum number $n$
from its purely Coulomb value is given by the matrix element~\cite{Gasser:2007zt}
\eq\label{eq:master}
E-E_n=\langle\Psi_n|\bar T(E)|\Psi_n\rangle\, .
\en
Here, $|\Psi_n\rangle$ denotes the bound solution of the Schr\"odinger equation for the Coulomb potential, corresponding to the discrete eigenvalue $E_n$, and
$\bar T(E)$ is the solution to Eq.~\eqref{eq:Tmatrix} with the Coulomb exchange in the initial and final states removed (see Fig.~\ref{fig:Tbar} for a diagramatical respresention).

Summing the geometric series for $\bar T(E)$, we obtain
\eq
\bar T(E)=\frac{4\pi A_{p\bar p}/m}{1-4\pi A_{p \bar p}/m\,\langle \bar g_C(E)\rangle}\, ,
\en
where the sum of Coulomb bubbles is
\eq
\langle \bar g_C(E)\rangle=\int\frac{d^3\bm{p}'}{(2\pi)^3}\,\frac{d^3\bm{p}}{(2\pi)^3}\,
\langle \bm{p}'|\bar g_C(E)|\bm{p}\rangle\, ,
\en
with
\eq
\langle \bm{p}'|\bar g_C(E)|\bm{p}\rangle=
\langle \bm{p}'|g_C(E)|\bm{p}\rangle
-\frac{\Psi_n(\bm{p}')\Psi_n(\bm{p})}{E-E_n}\, ,
\en
and
\eq
&&\hspace*{-.7cm}\langle \bm{p}'|g_C(E)|\bm{p}\rangle =\frac{(2\pi)^3\delta^{(3)}(\bm{p}'-\bm{p})}
{E-\bm{p}^2/m}
\\
&&\hspace*{-.7cm}~-\,\frac{m}{mE-{\bm{p}'}^2}\,\left[\frac{4\pi\alpha}{|\bm{p}'-\bm{p}|^2}\,
+I(E;\bm{p}',\bm{p})\right]
\frac{m}{mE-\bm{p}^2}\,.
\nonumber
\en
The integral $I$ is given by
\eq
I(E;\bm{p}',\bm{p})=4\pi\alpha\eta\int_0^1\frac{x^{-\eta}{dx}}{D}\,,
\en
where
\eq
D&=&(\bm{p}'-\bm{p})^2x
\nonumber\\
&+&\,\eta^2/\alpha^2
  (1-x)^2(E-{\bm{p}'}^2/m)(E-\bm{p}^2/m)\, ,
\en
and $\eta=\alpha(-m/E)^{1/2}/2$ is the Sommerfeld parameter. In our approximation, we are allowed to
replace $E$ by $E_n$ in the r.h.s. of Eq.~(\ref{eq:master}) because $E=E_n+O(\epsilon^3)$. As a result, we arrive at
\eq
\hspace*{-.5cm}\Delta E_n=E-E_n=|\tilde\Psi_n(0)|^2\frac{4\pi A_{p\bar p}/m}{1-4\pi A_{p\bar p}/m\,\langle \bar g_C(E_n)\rangle}\, ,
\en
where $\tilde \Psi_n(\bm{r})$ is the coordinate-space wave function of the unperturbed
level with the principal quantum number $n$. Both $\tilde\Psi_n(0)$ and
$\langle \bar g_C(E_n)\rangle$ are explicitly known. The latter contains an ultraviolet divergence that should be removed by the $O(e^2)$ renormalization of the strong non-derivative 
coupling in the initial Lagrangian. Since the issue is again very standard~\cite{Gasser:2007zt}, we quote here the final result only for the ground state with $n=1$
\eq\label{eq:DeltaE}
\Delta E_1=\frac{\frac{1}{2}\alpha^3m^2\tilde A_{p\bar p}}
{1-\alpha m\tilde A_{p\bar p}(\ln\alpha-1+i\pi)}\, .
\en
A similar formula for any $n\neq 1$ can be straightforwardly obtained.
Note that, as a result of renormalization, the quantity $A_{p\bar p}$ is replaced by $\tilde A_{p\bar p}$,
which is the Coulomb modified scattering length obtained from the $p\bar p\to p\bar p$ scattering amplitude, calculated in the
presence of photons~\cite{Gasser:2007zt}.\footnote{In the literature, there exist different definitions of the ``scattering length in the presence of Coulomb interactions'' and hence, this statement can potentially lead to confusion. For example, $\bar A_{p\bar p}$ is {\em not} the same scattering length that appears in the Trueman formula
\cite{Trueman:1961zza} (although these two are algebraically related at the lowest order in the fine structure constant). We stick to the definition given in Ref.~\cite{Gasser:2007zt}} The crucial point here is that $\tilde A_{p\bar p}-A_{p\bar p}=O(\epsilon)$, i.e., the potentially large logarithms are singled out (and summed up to all
orders) in Eq.~(\ref{eq:DeltaE}). At the order we are working, one can thus assume
$\tilde A_{p\bar p}=A_{p\bar p}$, and relegate all $O(\epsilon)$ corrections to the systematic error.\footnote{Strictly speaking, the term $-1$ in the parentheses in the denominator of Eq.~(\ref{eq:DeltaE}) also corresponds to the $O(\epsilon)$ correction and is kept for historical reasons only.}

To summarize, Eq.~(\ref{eq:DeltaE}) includes the leading $O(\epsilon^{1/2})$
and $O(\epsilon\ln\epsilon)$ corrections to the DGBT formula for the shift of the
ground-state energy level. It is obvious that no new parameters arise in this formula. The
sub-leading $O(\epsilon)$ corrections are no more parameter-free and are not
displayed in this equation. Furthermore, for all partial waves except the S-wave, no meaningful prediction beyond the leading-order DGBT formula can be done without performing a complex matching to isospin-violating effects in non-perturbative calculations with poorly known chiral $N\bar N$ interactions \cite{Dai:2017ont}.

It remains to be discussed how one can estimate the systematic error associated with the
$O(\epsilon)$ corrections. It is fair to say that we do not have any reliable information about these terms beyond simple dimensional arguments.
Since the expansion parameter $\epsilon\sim\alpha\sim (m_d-m_u)/\Lambda_{\rm had}$ is numerically about 1\%, we assign a rather conservative 5\,\% uncertainty to the input scattering lengths that translates into the error for the level shifts. 

{\bf 5.} In order to get the feeling of the size of the corrections, we have calculated
the singlet and triplet ground-state energies, using the scattering lengths
calculated from an antinucleon-nucleon interaction at next-to-next-to-next-to-leading order in chiral effective field theory~\cite{Dai:2017ont}. Our results are shown in Table~\ref{tab:table} and compared with the DGBT formula, Eq.~\eqref{eq:DeltaEDGBT} and the results of \cite{Dai:2017ont} based on the Trueman formula \cite{Trueman:1961zza}. 
Since the DGBT formula constitutes the leading order in
NREFT and the dominant corrections are $O(\epsilon^{1/2})\sim 0.1$, 
one may conclude that the uncertainty of the DGBT results should be of order of $10\%$. This estimate is roughly consistent with the numbers in Table~\ref{tab:table}.
We also compare with several experimental results measured in the 1990's at LEAR \cite{ASTERIX:1988rnq,  Heitlinger:1991cn,Augsburger:1999yt}.

\begin{table}[t]
  \begin{center}
    \begin{tabular}{|l|c|c|}
      \hline
      \multicolumn{3}{|c|}{Singlet}\\
      \hline
                         & $\mbox{Re}\,\Delta E_1$ [eV] & $\Gamma_1$ [eV]\\
      \hline\hline
      DGBT & $-368 $ & $1569$ \\
    \hline
      Total & $-476\pm 24$ & $1121\pm 44$\\
        w/o resummation & $-500\pm 27$ & $1087\pm 39$\\
  w/o cusp & $-408\pm 21$ & $1273\pm 51$\\
      \hline
      Dai {\it et al.} \cite{Dai:2017ont}& -443 & 1171\\
      \hline
      Augsburger {\it et al.} \cite{Augsburger:1999yt} &$-440 \pm 75$ & $1200\pm 250$ \\
      Ziegler {\it et al.} \cite{ASTERIX:1988rnq}&  $-740\pm 150$ & $1600 \pm 400$\\
      \hline
            \multicolumn{3}{c}{}\\
        \hline
         \multicolumn{3}{|c|}{Triplet}\\
      \hline
                         & $\mbox{Re}\,\Delta E_1$ [eV]& $\Gamma_1$ [eV]\\
      \hline\hline
      DGBT & $-806$ & $1595$ \\
      \hline
      Total & $-773\pm 34$ & $1106\pm 37$\\
      w/o resummation & $-789\pm 36$ & $955\pm 17$\\
  w/o cusp & $-721\pm 32$ & $1106\pm 37$\\
  \hline
      Dai {\it et al.} \cite{Dai:2017ont}& -770  & 1161  \\
      \hline
      Augsburger {\it et al.} \cite{Augsburger:1999yt} & $-785\pm 35$ & $940\pm 80$\\
      Heitlinger {\it et al.} \cite{Heitlinger:1991cn} & $-850\pm 42$ & $770\pm 150$\\
      \hline
    \end{tabular}
  \end{center}
  \caption{The results for the ground-state energy shift and width from Eq.~(\ref{eq:DeltaE}),
    using the strong scattering lengths from Ref.~\cite{Dai:2017ont}. 
    For comparison, we also show the level shifts of~\cite{Dai:2017ont} obtained by first calculating the Coulomb-corrected $p\bar p\to p\bar p$ scattering length and substituting it into the Trueman formula~\cite{Trueman:1961zza}. Moreover, we compare to the DGBT formula and experimental results from LEAR \cite{ASTERIX:1988rnq,Heitlinger:1991cn,Augsburger:1999yt}. Since the DGBT formula constitutes the leading order in NREFT, one may attach an uncertainty of about 10\% to the DGBT result from the size of the expansion parameter.
   }
\label{tab:table}
\end{table}
      
As one concludes from the table, the corrections to the leading-order DGBT formula are
very large and definitely have to be taken into account, in order to describe the experimental level shifts \cite{ASTERIX:1988rnq,Heitlinger:1991cn,Augsburger:1999yt}. 
Individual corrections -- both at
$O(\epsilon^{1/2})$ and $O(\epsilon\ln\epsilon)$ are also large (we remind the reader that they were not taken into account in the only available treatment of the problem within NREFT~\cite{Kong:1998ps}). Resummation of the
leading logarithmic terms $(\epsilon\ln\epsilon)^n$ leads to a rather significant effect. Furthermore, the results based on the use of the Trueman formula in~\cite{Dai:2017ont} lie in the same ballpark as ours, albeit they do not include the cusp effect in the amplitude and have no uncertainty estimates~\cite{private-Ulf}. We remind the reader that the leading-order Trueman formula properly accounts for the $\epsilon\ln\epsilon$ term through replacing $A_{p\bar p}$ by the Coulomb-modified scattering length. In order to do this, however, one has to solve the Lippmann-Schwinger equation in the presence of the Coulomb force numerically. In the NREFT approach one deals directly with the ``purely strong'' scattering amplitude $A_{p\bar p}$ and an analytic expression for the correction is available. One the one hand,
note that the errors that stem from a 5\,\% uncertainty
in the threshold amplitude are still moderate and will not hinder the extraction of the strong
amplitude $A_{p \bar p}$ from data at a reasonable accuracy. For a definitive test of the NREFT results based on the scattering lengths from \cite{Dai:2017ont}, on the other hand, more precise measurements are required.

{\bf 6.} In the past and present, the spectrum of the $\bar pp$ bound system has been extensively
studied within potential scattering theory, see, e.g.,~\cite{Buck:1977rt,Carbonell:1989cs,Carbonell:1992wd,Ydrefors:2021pmp}. Realistic phenomenological potentials have been used and different corrections to the DGBT formula have been identified. Without going into details, we note a fundamental
difference between our calculations and those carried within the potential model
that makes a direct, term-by-term comparison of the results meaningless. Namely, the
counting rules are lost in the potential model which can be viewed as a theory with
a smooth cutoff at  some momentum scale. 

We demonstrate this statement for the
simplest example. As discussed in {\bf 3.}, the cusp effect in the partial wave
with the orbital momentum $\ell$ scales as $O(\epsilon^{\ell+1/2})$ like the phase space.
In order to demonstrate this in dimensional regularization, one has to evaluate the neutron bubble at the $p\bar p$
threshold
\eq
\rho(E=2m)&=&\int\frac{d^d\bm{p}}{(2\pi)^d}\,\frac{\bm{p}^{2\ell}}{\bm{p}^2+2m(m_n-m)}
\nonumber\\[2mm]
&=&\delta^{2\ell+1}\frac{(-1)^{\ell+1}}{4\pi}=O(\epsilon^{\ell+1/2})\, .
\label{eq:threshold}
\en
Therefore this effect must be very small in higher partial waves.

Now, for simplicity, assume that the strong potential has the separable form
with the form factor $v(\bm{p})={\Lambda^2}/{(\Lambda^2+\bm{p}^2)}$ (a different
choice of the potential will not change the result qualitatively). Then, the bubble
takes the form
\eq
\rho_\Lambda(E=2m)=\int\frac{d^3\bm{p}}{(2\pi)^3}\,\frac{\bm{p}^{2\ell}v^2(\bm{p})}{\bm{p}^2+2m(m_n-m)}\, .
\en
Expansion at small $\delta$ gives
\eq
&&\hspace*{-1.3cm}\ell=0:~ \rho_\Lambda(E=2m)=\frac{\Lambda}{8\pi}-\frac{\delta}{4\pi}+O(\delta^2)\, ,
\nonumber\\
&&\hspace*{-1.3cm}\ell=1:~ \rho_\Lambda(E=2m)=\frac{\Lambda^3}{8\pi}-\frac{\Lambda\delta^2}{8\pi}+\frac{\delta^3}{4\pi}+O(\delta^4)\, ,
\en
where $\delta$ is defined in Eq.~\eqref{eq:deltadef}
and so on. The $\delta$-independent terms merely modify the isospin-symmetric
scattering amplitude and will not be considered in the following. One sees that in the S-wave, the term linear in $\delta$ (i.e., the non-analytic order $\epsilon^{1/2}$ piece) corresponds to the leading isospin-breaking correction, whereas the analytic terms containing even powers of $\delta$ are subleading. The situation is reversed in higher partial waves, where corrections always start at order $\delta^2$ (order $\epsilon$), and the physically significant non-analytic
correction appears at order $\delta^{2\ell+1}$ first. In other words, in all partial waves
other than the S-wave, the cusp correction is buried behind the structure-dependent analytic terms which are generally not known (the same happens in the case of Coulomb corrections as well). One could also conclude that the results of calculations with two different
potentials should be always compatible within the systematic error, provided the potentials do not feature unnatural scales.

{\bf 7.} In conclusion, we have studied the strong level shifts in the protonium atom using the well-established framework of non-relativistic effective field theory (NREFT). This 
theory has successfully been applied to a variety of atoms, namely
$\pi\pi$, $\pi K$, $\pi H$, $\pi d$, $K H$, and $K d$ atoms, before \cite{Gasser:2007zt}. The order $\epsilon$ and $\epsilon \ln \epsilon$ corrections to the 
DGBT formula are sizable and will significantly affect the extraction of the $N\bar N$
scattering lengths from level shift data for protonium. Using $N\bar N$
scattering lengths from chiral EFT as input, the S-wave level shifts are consistent with previous experiments \cite{ASTERIX:1988rnq,Heitlinger:1991cn,Augsburger:1999yt}. The experimental uncertainties are comparable with the theory uncertainties for the triplet state but a factor three larger for the singlet states. Once more precise experiments become available, it will be necessary to calculate higher-order structure-dependent corrections. It remains to be seen whether a reliable calculation of these corrections in NREFT is possible. Such a calculation is beyond the scope of the present study. For higher partial waves, the structure-dependent corrections enter already at leading order. 

Of course, applying the same method to study
of heavier systems is a much more challenging enterprise. In addition to the measurement of  the neutron skin of heavy nuclei in PUMA \cite{PUMA:2022ngr}, it could unveil interesting information about the antiproton interactions with
light and heavy nuclei. The present paper is only a first, small step in this direction.


\section*{Acknowledgements} The authors thank Ulf-G. Mei{\ss}ner and Alexandre Obertelli
for useful discussions and Pierre Kn\"otzele for comments on the manuscript.
The work of A.R. was  funded in part by Deutsche Forschungsgemeinschaft
(DFG, German Research Foundation)  -- Project number RU 1205/2-1,
 and the Ministry of Culture and Science of North Rhine-Westphalia through the NRW-FAIR project. H.-W.H. was supported in part by the Deutsche Forschungsgemeinschaft (DFG, German Research Foundation) - Project ID 279384907 - SFB 1245 and by the BMFTR Contract No. 05P24RDB.

\bibliographystyle{elsarticle-num}
\bibliography{ref1}

\begin{thebibliography}{10}
\expandafter\ifx\csname url\endcsname\relax
  \def\url#1{\texttt{#1}}\fi
\expandafter\ifx\csname urlprefix\endcsname\relax\def\urlprefix{URL }\fi
\expandafter\ifx\csname href\endcsname\relax
  \def\href#1#2{#2} \def\path#1{#1}\fi

\bibitem{PUMA:2022ngr}
T.~Aumann, et~al., {PUMA, antiProton unstable matter annihilation}, Eur. Phys.
  J. A 58~(5) (2022) 88.
\newblock \href {https://doi.org/10.1140/epja/s10050-022-00713-x}
  {\path{doi:10.1140/epja/s10050-022-00713-x}}.

\bibitem{Desai:1960zz}
B.~R. Desai, {Proton-Antiproton Annihilation in Protonium}, Phys. Rev. 119
  (1960) 1385--1389.
\newblock \href {https://doi.org/10.1103/PhysRev.119.1385}
  {\path{doi:10.1103/PhysRev.119.1385}}.

\bibitem{Buck:1977rt}
W.~W. Buck, C.~B. Dover, J.~M. Richard, {The Interaction of Nucleons with
  anti-Nucleons. 1. General Features of the anti-N n Spectrum in Potential
  Models}, Annals Phys. 121 (1979) 47.
\newblock \href {https://doi.org/10.1016/0003-4916(79)90091-5}
  {\path{doi:10.1016/0003-4916(79)90091-5}}.

\bibitem{Carbonell:1989cs}
J.~Carbonell, G.~Ihle, J.~M. Richard, {Protonium Annihilation in Optical
  Models}, Z. Phys. A 334 (1989) 329--341.

\bibitem{Carbonell:1992wd}
J.~Carbonell, J.-M. Richard, S.~Wycech, {On the relation between protonium
  level shifts and nucleon - anti-nucleon scattering amplitudes}, Z. Phys. A
  343 (1992) 325--329.

\bibitem{Vandermeulen:1992eh}
J.~Vandermeulen, {Protonium decay and phase space}, Z. Phys. A 342 (1992)
  329--341.
\newblock \href {https://doi.org/10.1007/BF01291517}
  {\path{doi:10.1007/BF01291517}}.

\bibitem{Kong:1998ps}
X.~Kong, F.~Ravndal, {Higher order hadronic energy level shifts in protonium
  from effective field theory} (1998).
\newblock \href {http://arxiv.org/abs/nucl-th/9803046}
  {\path{arXiv:nucl-th/9803046}}.

\bibitem{Klempt:2002ap}
E.~Klempt, F.~Bradamante, A.~Martin, J.~M. Richard, {Antinucleon nucleon
  interaction at low energy: Scattering and protonium}, Phys. Rept. 368 (2002)
  119--316.
\newblock \href {https://doi.org/10.1016/S0370-1573(02)00144-8}
  {\path{doi:10.1016/S0370-1573(02)00144-8}}.

\bibitem{Ydrefors:2021pmp}
E.~Ydrefors, J.~Carbonell, {Protonium annihilation densities in a unitary
  coupled channel model}, Eur. Phys. J. A 57~(11) (2021) 303.
\newblock \href {http://arxiv.org/abs/2110.08628} {\path{arXiv:2110.08628}},
  \href {https://doi.org/10.1140/epja/s10050-021-00609-2}
  {\path{doi:10.1140/epja/s10050-021-00609-2}}.

\bibitem{Cabrera-Trujillo:2023ify}
R.~Cabrera-Trujillo, C.~E. d. l.~C. Roman, C.~E. Teran-Cisneros, {Ionization,
  excitation, protonium formation, and energy loss of antiprotons colliding
  with atomic hydrogen}, Phys. Rev. A 108~(1) (2023) 012817.
\newblock \href {https://doi.org/10.1103/PhysRevA.108.012817}
  {\path{doi:10.1103/PhysRevA.108.012817}}.

\bibitem{Ma:2024gsw}
B.-Q. Ma, {Protonium: Discovery and prediction}, Chin. Sci. Bull. 69 (2024)
  4620.
\newblock \href {http://arxiv.org/abs/2406.19180} {\path{arXiv:2406.19180}},
  \href {https://doi.org/10.1360/TB-2024-0578}
  {\path{doi:10.1360/TB-2024-0578}}.

\bibitem{Gasser:2007zt}
J.~Gasser, V.~E. Lyubovitskij, A.~Rusetsky, {Hadronic atoms in QCD + QED},
  Phys. Rept. 456 (2008) 167--251.
\newblock \href {http://arxiv.org/abs/0711.3522} {\path{arXiv:0711.3522}},
  \href {https://doi.org/10.1016/j.physrep.2007.09.006}
  {\path{doi:10.1016/j.physrep.2007.09.006}}.

\bibitem{Gasser:2009wf}
J.~Gasser, V.~E. Lyubovitskij, A.~Rusetsky, {Hadronic Atoms}, Ann. Rev. Nucl.
  Part. Sci. 59 (2009) 169--190.
\newblock \href {http://arxiv.org/abs/0903.0257} {\path{arXiv:0903.0257}},
  \href {https://doi.org/10.1146/annurev.nucl.010909.083806}
  {\path{doi:10.1146/annurev.nucl.010909.083806}}.

\bibitem{Epelbaum:2004fk}
E.~Epelbaum, W.~Gl{\"o}ckle, U.-G. Mei{\ss}ner, {The Two-nucleon system at
  next-to-next-to-next-to-leading order}, Nucl. Phys. A 747 (2005) 362--424.
\newblock \href {http://arxiv.org/abs/nucl-th/0405048}
  {\path{arXiv:nucl-th/0405048}}, \href
  {https://doi.org/10.1016/j.nuclphysa.2004.09.107}
  {\path{doi:10.1016/j.nuclphysa.2004.09.107}}.

\bibitem{Deser:1954vq}
S.~Deser, M.~L. Goldberger, K.~Baumann, W.~E. Thirring, {Energy level
  displacements in pi mesonic atoms}, Phys. Rev. 96 (1954) 774--776.
\newblock \href {https://doi.org/10.1103/PhysRev.96.774}
  {\path{doi:10.1103/PhysRev.96.774}}.

\bibitem{Gasser:2003hk}
J.~Gasser, A.~Rusetsky, I.~Scimemi, {Electromagnetic corrections in hadronic
  processes}, Eur. Phys. J. C 32 (2003) 97--114.
\newblock \href {http://arxiv.org/abs/hep-ph/0305260}
  {\path{arXiv:hep-ph/0305260}}, \href
  {https://doi.org/10.1140/epjc/s2003-01383-1}
  {\path{doi:10.1140/epjc/s2003-01383-1}}.

\bibitem{Gasser:1999vf}
J.~Gasser, V.~E. Lyubovitskij, A.~Rusetsky, {Numerical analysis of the $\pi^+
  \pi^-$ atom lifetime in ChPT}, Phys. Lett. B 471 (1999) 244--250.
\newblock \href {http://arxiv.org/abs/hep-ph/9910438}
  {\path{arXiv:hep-ph/9910438}}, \href
  {https://doi.org/10.1016/S0370-2693(99)01334-9}
  {\path{doi:10.1016/S0370-2693(99)01334-9}}.

\bibitem{Gall:1999bn}
A.~Gall, J.~Gasser, V.~E. Lyubovitskij, A.~Rusetsky, {On the lifetime of the
  $\pi^+ \pi^-$ atom}, Phys. Lett. B 462 (1999) 335--340.
\newblock \href {http://arxiv.org/abs/hep-ph/9905309}
  {\path{arXiv:hep-ph/9905309}}, \href
  {https://doi.org/10.1016/S0370-2693(99)00918-1}
  {\path{doi:10.1016/S0370-2693(99)00918-1}}.

\bibitem{Kubis:2001ij}
B.~Kubis, U.-G. Mei{\ss}ner, {Isospin violation in pion kaon scattering}, Nucl.
  Phys. A 699 (2002) 709--731.
\newblock \href {http://arxiv.org/abs/hep-ph/0107199}
  {\path{arXiv:hep-ph/0107199}}, \href
  {https://doi.org/10.1016/S0375-9474(01)01318-5}
  {\path{doi:10.1016/S0375-9474(01)01318-5}}.

\bibitem{Kubis:2001bx}
B.~Kubis, U.-G. Mei{\ss}ner, {Isospin violation in low-energy charged pion kaon
  scattering}, Phys. Lett. B 529 (2002) 69--76.
\newblock \href {http://arxiv.org/abs/hep-ph/0112154}
  {\path{arXiv:hep-ph/0112154}}, \href
  {https://doi.org/10.1016/S0370-2693(02)01192-9}
  {\path{doi:10.1016/S0370-2693(02)01192-9}}.

\bibitem{Gasser:2002am}
J.~Gasser, M.~A. Ivanov, E.~Lipartia, M.~Mojzis, A.~Rusetsky, {Ground state
  energy of pionic hydrogen to one loop}, Eur. Phys. J. C 26 (2002) 13--34.
\newblock \href {http://arxiv.org/abs/hep-ph/0206068}
  {\path{arXiv:hep-ph/0206068}}, \href
  {https://doi.org/10.1007/s10052-002-1013-z}
  {\path{doi:10.1007/s10052-002-1013-z}}.

\bibitem{Dai:2017ont}
L.-Y. Dai, J.~Haidenbauer, U.-G. Mei{\ss}ner, {Antinucleon-nucleon interaction
  at next-to-next-to-next-to-leading order in chiral effective field theory},
  JHEP 07 (2017) 078.
\newblock \href {http://arxiv.org/abs/1702.02065} {\path{arXiv:1702.02065}},
  \href {https://doi.org/10.1007/JHEP07(2017)078}
  {\path{doi:10.1007/JHEP07(2017)078}}.

\bibitem{Trueman:1961zza}
T.~L. Trueman, {Energy level shifts in atomic states of strongly-interacting
  particles}, Nucl. Phys. 26 (1961) 57--67.
\newblock \href {https://doi.org/10.1016/0029-5582(61)90115-8}
  {\path{doi:10.1016/0029-5582(61)90115-8}}.

\bibitem{Meissner:2004jr}
U.~G. Mei{\ss}ner, U.~Raha, A.~Rusetsky, {Spectrum and decays of kaonic
  hydrogen}, Eur. Phys. J. C 35 (2004) 349--357.
\newblock \href {http://arxiv.org/abs/hep-ph/0402261}
  {\path{arXiv:hep-ph/0402261}}, \href
  {https://doi.org/10.1140/epjc/s2004-01859-4}
  {\path{doi:10.1140/epjc/s2004-01859-4}}.

\bibitem{Meissner:2006gx}
U.-G. Mei{\ss}ner, U.~Raha, A.~Rusetsky, {Kaon-nucleon scattering lengths from
  kaonic deuterium experiments}, Eur. Phys. J. C 47 (2006) 473--480.
\newblock \href {http://arxiv.org/abs/nucl-th/0603029}
  {\path{arXiv:nucl-th/0603029}}, \href
  {https://doi.org/10.1140/epjc/s2006-02578-6}
  {\path{doi:10.1140/epjc/s2006-02578-6}}.

\bibitem{ASTERIX:1988rnq}
M.~Ziegler, et~al., {Measurement of the Strong Interaction Shift and Broadening
  of the Ground State of the $p \bar{p}$ Atom}, Phys. Lett. B 206 (1988)
  151--158.
\newblock \href {https://doi.org/10.1016/0370-2693(88)91279-8}
  {\path{doi:10.1016/0370-2693(88)91279-8}}.

\bibitem{Heitlinger:1991cn}
K.~Heitlinger, R.~Bacher, A.~Badertscher, P.~Bl{\"u}m, J.~Eades, J.~Egger,
  K.~Elsener, D.~Gotta, E.~Morenzoni, L.~M. Simons, {Precision measurement of
  anti-protonic hydrogen and deuterium x-rays}, Z. Phys. A 342 (1992) 359--368.
\newblock \href {https://doi.org/10.1007/BF01291519}
  {\path{doi:10.1007/BF01291519}}.

\bibitem{Augsburger:1999yt}
M.~Augsburger, et~al., {Measurement of the strong interaction parameters in
  anti-protonic hydrogen and probable evidence for an interference with inner
  bremsstrahlung}, Nucl. Phys. A 658 (1999) 149--162.
\newblock \href {https://doi.org/10.1016/S0375-9474(99)00352-8}
  {\path{doi:10.1016/S0375-9474(99)00352-8}}.

\bibitem{private-Ulf}
U.-G. Mei{\ss}ner, private communication.

\end{thebibliography}

\end{document}